\documentclass[aps,pra,onecolumn,longbibliography,superscriptaddress,final,floatfix, 11pt,tightenlines]{revtex4-2}

\usepackage{amsfonts}
\usepackage{amsmath}
\usepackage{amssymb}
\usepackage{subfigure}
\usepackage{latexsym}
\usepackage[export]{adjustbox}
\usepackage{array}
\newcolumntype{P}[1]{>{\centering\arraybackslash}p{#1}}
\newcolumntype{M}[1]{>{\centering\arraybackslash}m{#1}}
\usepackage{caption}
\usepackage[colorlinks,citecolor=red,urlcolor=blue,bookmarks=false,hypertexnames=true]{hyperref} 
\usepackage[usenames, dvipsnames]{color}
\usepackage[svgnames]{xcolor}
\usepackage{color}
\usepackage{bm}
\usepackage{relsize}
\usepackage{float}
\usepackage{subfloat}
\usepackage{dcolumn}
\usepackage{epsfig}
\usepackage{graphicx}
\usepackage{multirow}
\usepackage{makecell}
\usepackage{mathtools}

\usepackage{footnote}
\usepackage{balance}

\usepackage[mathscr]{euscript}

\DeclareSymbolFont{rsfs}{U}{rsfs}{m}{n}
\DeclareSymbolFontAlphabet{\mathscrsfs}{rsfs}

\begin{document}

\title{Decoherence of a charged Brownian particle in a magnetic field : an analysis of the roles of coupling via position and momentum variables }
\author{Suraka Bhattacharjee}
\affiliation{SASTRA Deemed University, Tirumalaisamudram, Thanjavur-613401, India} 
\author{Koushik Mandal}
\affiliation{Department of Applied Science, Haldia Institute of Technology, Haldia, 721657, India}    
\author{Supurna Sinha}
\affiliation{Raman Research Institute, Bangalore-560080, India}
\date{\today}
\begin{abstract}
 The study of decoherence plays a key role in our understanding of the transition from the quantum to the classical world. 
 Typically, one considers a system coupled to an external bath which forms a model for an open quantum system.
 While most of the studies pertain to a position coupling between the system and the environment, 
 some involve a momentum coupling, 
 giving rise to an anomalous diffusive model. 
 Here we have gone beyond existing studies and analysed the quantum Langevin dynamics of a harmonically oscillating charged Brownian particle in the presence of a magnetic field and coupled to an Ohmic heat bath via both position and momentum couplings. The presence of both position and momentum couplings leads to a stronger interaction with the environment, resulting in a faster loss of coherence compared to a situation where only position coupling is present. The rate of decoherence can be tuned by
 controlling the relative strengths of the position and momentum coupling parameters. In addition, the magnetic field results in the slowing down of the loss of information from the system, irrespective of the nature of coupling between the system and the bath.
 Our results can be experimentally verified
by designing a suitable ion trap setup.
 
\end{abstract}
\maketitle
\section{Introduction}
Decoherence plays a crucial role in understanding the transition from the quantum to the classical domain \cite{Zeh}. An isolated quantum system can exist in a coherent superposition of a number of possible states. A coupling between such a quantum system and an environment leads to destruction of coherence and emergence of classical probabilities 
where only one of various possible states can occur. This is the process of environment induced decoherence  \cite{feynman2000theory,Caldeira1}.
The theoretical ideas related to environment-induced decoherence date back to the sixties \cite{feynman2000theory,Zeh}. In recent years, it has turned out 
to be of great technological relevance as it poses a major obstacle to implementation of large-scale quantum computing \cite{Ollerenshow, Duan}.
The process of decoherence has been analysed to study various aspects of open quantum systems \cite{Zeh}.
Many of the studies involve a Brownian particle coupled to an external heat bath \cite{Caldeira1}. The random motion of a charged particle coupled to a heat bath has been studied in detail via the Langevin equation which has both deterministic and stochastic forces and captures the effect of interaction between the system and the environment \cite{ford,Ford1}. In \cite{Ghorashi}, the two dimensional master equation for a Brownian particle in the presence of a magnetic field on a non-commutative plane was derived and decoherence was investigated via the rate of linear entropy in a non-commutative plane. Further, the effect of electric field on the decoherence of a charged Brownian particle was studied in terms of the non-Markovian master equation determined by two coupled Green’s functions \cite{Ghorashi}. Apart from its fundamental relevance to understanding the issue of quantum to classical transition, decoherence is of great experimental relevance. In recent years the effect of decoherence has been reported in cavity quantum electrodynamics (QED) \cite{PhysRevLett.77.4887}, qubit-superconducting systems \cite{burnett2019decoherence,burkard2004multilevel}, matter-wave interferometry \cite{xu2010experimental}, and ion traps \cite{schneider1998decoherence,schneider1999decoherence}. In addition, 
decoherence has been experimentally detected in quantum dots \cite{qtm_dot_deco}, quantum mechanical resonators \cite{RevModPhys.86.1391} and in Bose-Einstein condensates (BEC) \cite{kuang1999decoherence}. \\
In one of our earlier works \cite{bhattacharjee2023decoherence}, we have analyzed decoherence for a charged harmonically oscillating Brownian particle in the presence of a 
magnetic field 
and coupled to an external heat bath characterised by  Ohmic and non-Ohmic spectral densities. In \cite{bhattacharjee2023decoherence}
we had considered a coupling via position coordinates between the system and bath. There have been studies of anomalous dissipative models, involving a momentum coupling between the system and heat bath \cite{cuccoli,huang2022exact,Bai,Ankerhold,Ferialdi}. The generalized Caldeira Leggett (GCL) Model was used in \cite{Bai}, where all types of position and momentum couplings were considered to derive the mean squared velocity and the velocity correlation functions corresponding to an anomalous diffusive model. Furthermore, the non-Markovian  master equation was derived and analysed for position-momentum coupling and the enhancement of dissipation in the presence of momentum coupling was highlighted \cite{Ferialdi}. In \cite{Leggett} the study of anomalous dissipation has shown the emergence of a `pseudo Langevin'  equation where the damping constant involves a second derivative of the potential term. For the harmonic potential the quantum Langevin equation is reduced to the standard form, but the damping coefficient retains system dependent terms apart from bath dependent terms obtained in the case of normal diffusion \cite{Leggett}. This type of dissipation can be physically observed when the effect of electromagnetic black body radiation on Josephson junction is considered \cite{ferrero2012new,cuccoli}. Moreover, the concept of `mixed diffusion' was also put forward by the author in \cite{Leggett}, in the context of a liquid-filled cylinder suspended by a highly non-linear torsion thread, where the diffusion appears to be normal or anomalous depending on the `collision-less' or `hydrodynamic' limits of the equation of motion respectively. There have been subsequent studies of anomalous diffusion and the non-Markovian dynamics of a Brownian particle coupled to a heat bath through position-momentum coupling has been analysed within the framework of anomalous dissipation \cite{Ferialdi}.  In this paper, we have used a modified form of the generalized Caldeira Leggett model used in \cite{Bai}, where only the position-position and the momentum-momentum coupling terms are retained. Applying this anomalous diffusion model, we have derived the quantum Langevin equation and the noise correlation for a Brownian particle in the presence of a magnetic field and coupled to an Ohmic heat bath. In addition, we have made use of the noise correlation and analysed the decoherence for this model, within the framework of non-Markovian dynamics, which ensures that the environment preserves the memory of its collision with the Brownian particle. However, we have assumed the validity of the Born approximation in our analysis \cite{Schlosshauer}. Further, we have extended our analysis to highlight the effect of the magnetic field on the system which plays a significant role in shaping the rate of decoherence and the Quantum-Classical transition in the open quantum system.\\
The paper is arranged as follows: in Section-II we have formulated the quantum Langevin equation for a harmonically oscillating charged Brownian particle in the presence of a magnetic field and coupled to a heat bath through position and momentum coordinate couplings. The quantum Langevin equation pertains to an Ohmic heat bath with an abrupt cut-off model for the spectral density of the bath. In Section-III we have derived a non-Markovian master equation for a charged particle coupled to a heat bath and decoherence for the open quantum system has been studied in Section-IV. The numerical results and the plots are shown in Section-V and the paper is concluded with some discussions and concluding remarks in Section-VI. 
\section{Formulation of the quantum Langevin Equation}
The Hamiltonian for a charged Brownian particle harmonically oscillating in the presence of a magnetic field and coupled to an external heat bath through position and momentum coordinate couplings is given by:
\begin{align}
  H=\frac{1}{2m}\left(\vec{p}-\frac{q\vec{A}}{c} \right)^2+\frac{1}{2}m\omega_0^2 r^2+\sum_{j=1}^N\left[\frac{1}{2m_j}\left(\vec{p_j}-g_j\vec{p}+\frac{g_j q}{c} \vec{A} \right)^2 +\frac{1}{2}m_j \omega_j^2 \left(\vec{q_j}-d_j\vec{r} \right)^2 \right] \label{Hamiltonian}
\end{align}
Here, $m$, $q$, $\vec{r}$, $\vec{p}$ are the mass, charge, position and momentum coordinates of the Brownian particle respectively and  $\vec{q_j}$, $\vec{p_j}$, $g_j$ and $d_j$ represent the position coordinate, momentum coordinate, strength of the coupling between the momentum and position coordinates of the particle and the momentum and the position coordinates of the $j^{th}$ bath oscillator respectively. $\vec{A}(r)$ represents the vector potential pertaining to the 
applied magnetic field $\vec{B}$ and $\omega_0$ is the harmonic oscillator frequency of the system particle.\\
As expected, Eq.(\ref{Hamiltonian}) reduces to the position coupling Hamiltonian in the limit of the coupling strength $g_{j}$ between the momentum variables going to zero and we recover the momentum coupling Hamiltonian for $d_j=0$. The quantum Langevin dynamics of a charged particle in a magnetic field and coupled to the bath via momentum variables has been reported earlier in the literature \cite{gupta2011quantum,bhattacharjee2022quantum}. In this work, we go beyond earlier studies and introduce both position and momentum system-bath couplings, study the Langevin dynamics and then analyze the process of decoherence.\\
We derive the velocity and the momentum of the Brownian particle by applying the Heisenberg equations of motion :
\begin{align}
    \vec{v}=\dot{\vec{r}}=\frac{1}{i\hbar}\left[\vec{r},H \right]=\frac{1}{m}\left(\vec{p}-\frac{q\vec{A}}{c}\right)-\sum_{j=1}^N\frac{g_j}{m_j}\left(\vec{p_j}-g_j \vec{p}+\frac{g_j q}{c}\vec{A}  \right) \label{rdot}
\end{align}
and in a similar way the equation for the $\alpha$-th component of the particle momentum appears as
\begin{align}
  \dot{p}_\alpha=&\frac{1}{i\hbar}[p_{\alpha},H]=\frac{1}{2m i \hbar}\left[p_\alpha,\left(\vec{p}-\frac{q \vec{A}}{c} \right)^2 \right] -m \omega_0^2 r_\alpha+\sum_j m_j \omega_j^2 d_j\left(q_{j \alpha}-d_j r_\alpha \right)
  \end{align}
Hence, the equation for the particle momentum in vector form can be written as
\begin{align}
  \dot{\vec{p}}=&\frac{q}{c}(\vec{v}\times \vec{B})+\frac{q}{c}(\vec{v}. \vec{\nabla}) \vec{A}+\frac{i \hbar q}{2 m_r c} \vec{\nabla}(\vec{\nabla}.\vec{A})- m\omega_0^2 \vec{r}+\sum_j m_j \omega_j^2 d_j(\vec{q_j}-d_j \vec{r}) \label{pdot}
\end{align}
where, $\alpha$ denotes the $x$ and $y$ components of the position and momentum coordinates $\vec{r}$ and $\vec{p}$.\\
Then we use the Heisenberg equation for the vector potential $\vec{A}$ as a function of the position coordinate $\vec{r}$.
\begin{align}
  \dot{A}_\alpha &=\frac{1}{i \hbar} \left[A_\alpha,H \right] = v_\beta (\partial_\beta A_\alpha)+\frac{i \hbar}{2 m_r}\partial_\beta \partial_\beta A_\alpha\\
  & \dot{\vec{A}}(r)= (\vec{v}.\vec{\nabla}) \vec{A}+\frac{i \hbar}{2 m_r} \nabla^2 \vec{A} \label{Adot}
\end{align}
From Eq.(\ref{rdot}) 
we get the time derivative of the particle momentum as:
\begin{align}
 \dot{\vec{p}}=& m_r \ddot{\vec{r}}+\frac{q}{c}\dot{\vec{A}}+ \sum_{j=1}^N \frac{g_j m_r}{m_j}\dot{\vec{p}}_j \label{pdotone}
\end{align}
where $m_r$ is the renormalized mass defined by:
\begin{align}
    m_r=m/\left[1+\sum_{j=1}^N \frac{g_j^2 m}{m_j}\right]
\end{align}
We now eliminate the momentum variables from Eqs.(\ref{pdot}) and (\ref{pdotone}) by replacing  $\dot{\vec{A}}$ in Eq.(\ref{pdotone}) by Eq.(\ref{Adot}) 
and equating Eqs.(\ref{pdot}) and (\ref{pdotone}):
\begin{align}
m_r \ddot{\vec{r}}=-m \omega_0^2 \vec{r}+\frac{q}{c}(\vec{v}\times \vec{B})+\sum_{j=1}^N g_j m_r \omega_j^2 \vec{q}_j + \sum_{j=1}^N m_j \omega_j^2 d_j(\vec{q_j}-d_j \vec{r})-\sum_{j=1}^N g_j m_r \omega_j^2 d_j \vec{r}   \label{Langevin1}
\end{align}
where,
\begin{align}
    \dot{p_j}=-m_j \omega_j^2 \left(\vec{q}_j-d_j \vec{r}  \right)
\end{align}
We then derive the quantum Langevin equation by using the retarded solution of the bath co-ordinates $q_j$ and by plugging it into Eq.(\ref{Langevin1}).\\
The equations of motion for the bath co-ordinates are given by:
\begin{align}
    \dot{\vec{q_j}}&=\frac{1}{i \hbar} \left[\vec{q_j},H\right]\\
    &= \frac{1}{m_j} \left(\vec{p_j}-g_j \vec{p} + \frac{g_j q \vec{A}}{c}\right) \label{qjdot}
\end{align}
  \begin{align}
      \dot{\vec{p_j}}=-m_j \omega_j^2 \left( \vec{q_j}-d_j \vec{r}\right) \label{pjdot}
  \end{align}  
From Eqns. (\ref{qjdot}) and (\ref{pjdot}) we derive the time evolution equation for the position coordinate as:
\begin{align}
    m_j \ddot{\vec{q_j}}&=\left( \dot{\vec{p_j}}-g_j \dot{\vec{p}} + \frac{g_j q}{c} \dot{\vec{A}} \right) \label{qjddot}
\end{align}
Using Eqs.(\ref{pdot}), (\ref{Adot}) and (\ref{pjdot})in Eq.(\ref{qjddot}), we get:
\begin{align}
   m_j \ddot{\vec{q_j}}=-m_j \omega_j^2  \left( \vec{q_j}-d_j \vec{r} \right)-\frac{g_j q}{c}\left(  \vec{v} \times \vec{B}\right)+g_j m \omega_0^2 \vec{r} -g_j \sum_{l=1} ^{N}m_l \omega_l^2 d_l
 \left(\vec{q_l}-d_l \vec{r}  \right)
 \end{align}\\
 \begin{align}
   m_j \ddot{\vec{q_j}}=-m_j \omega_j^2  \left( \vec{q_j}-d_j \vec{r} \right)-\frac{g_j q}{c}\left(  \vec{v} \times \vec{B}\right)+g_j m \omega_0^2 \vec{r} + K_j \vec{r} + \vec{Q}_j(t) \label{bathoscillatoreq}
 \end{align}
 where, $K_j$ and $\vec{Q}_j$ represent the weighted sum of the spring constants and the weighted sum of the harmonic forces acting on the bath oscillators individually oscillating with frequency $\omega_j$ respectively: 
 \begin{align}
     K_j=g_j \sum_{l=1}^{N} m_l \omega_l^2 d_l^2
 \end{align}
 \begin{align}
\vec{Q}_j(t)=-g_j \sum_{l=1}^{N}d_l m_l \omega_l^2 
 \vec{q_l}(t)
 \end{align}
 The retarded solution for $\vec{q_j}$ is given by:
\begin{align}
    \vec{q}_j(t)&=\vec{q}_j^h (t)+\frac{1}{m_j\omega_j^2}\int_0^t dt' \vec{Q}(t')\sin\left[\omega_j \left(t-t'\right) \right]+\frac{g_j m \omega_0^2 +d_j m_j \omega_j^2 +K_j}{m_j \omega_j^2} \left[\vec{r}(t)-\vec{r}(0) \cos \left(\omega_j t\right) \right]- \nonumber\\
    &\frac{g_j m \omega_0^2 +d_j m_j \omega_j^2 +K_j}{m_j \omega_j^2} \int_0^t dt' \dot{\vec{r}}(t')\cos \left[\omega_j \left(t-t' \right)\right] - \frac{g_j e \omega_j}{c  m_j \omega_j^2} \Gamma \int_0^t dt' \dot{\vec{r}}(t')\sin \left[\omega_j \left(t-t'\right)  \right] \label{qjretarded}       
    \end{align}
    where $\vec{q}_j^h (t)$ denotes the solution of the homogeneous equation corresponding to the time evolution equation for the bath oscillators (Eq.(\ref{bathoscillatoreq})).\\
    Inserting Eq.(\ref{qjretarded}) in Eq.(\ref{Langevin1}) we get:
    \begin{align}
        m_r \ddot{\vec{r}}-\frac{q}{c}\left(\vec{v}\times \vec{B} \right)+\int_0^t dt' \dot{\vec{r}}(t')\mu \left(t-t'\right) +\mu_d(t)\vec{r}(0)+m_1 \omega_0^2 \vec{r}=\vec{\mathcal{F}}_1(t)+\vec{\mathcal{F}}_2(t) \label{Langevinfinal}
    \end{align}
    where $\vec{\mathcal{F}}_1(t)$ and $\vec{\mathcal{F}}_2(t)$ are random forces pertaining to quantum noise.
    The memory function $\mu(t)$ has both diagonal and off-diagonal terms. $\mathcal{F}_1(t)$, $
    \mathcal{F}_2(t)$ and $\mu(t)$ are given by the following expressions:
    \begin{align}
        &\vec{\mathcal{F}}_1(t)=\sum_{j=1}^N \left( g_j m_r +d_j m_j  \right) \omega_j^2 \vec{q}_j^h (t)\Theta(t-t') \label{F1(t)}\\
        &\vec{\mathcal{F}}_2(t)=\sum_{j=1}^N \frac{\left( g_j m_r +d_j m_j \right)} {m_j\omega_j^2}\int_0^t dt' \vec{Q}(t')\sin\left[\omega_j \left(t-t'\right) \right]\Theta(t-t')\label{F2(t)}\\
        &\mu(t-t')=\mu_d(t-t')+\mu_{od}(t-t')\\
        &\mu_d(t-t')=\sum_{j=1}^N \left( g_j m_r + d_j m_j \right)  \frac{g_j m \omega_0^2+d_j m_j \omega_j^2+K_j}{m_j} \cos\left[\omega_j (t-t')  \right] \Theta (t-t') \\
        &\mu_{od}(t-t')=\sum_{j=1}^N \frac{-g_j e \Gamma \left( g_j m_r \omega_j^2+d_j m_j \omega_j^2 \right)}{cm\omega_j}  \sin \left[ \omega_j (t-t') \right]\Theta(t-t')
         \end{align}
         and $m_{1}$ in Eq.[\ref{Langevinfinal}] is
         \begin{align}
        m_1=&-\sum_{j=1}^N \left[ \left( \frac{g_j m \omega_0^2+d_j m_j \omega_j^2+K_j}{m_j \omega_j^2}+d_j \right) \left(g_j m_r +m_j d_j \right)\right] \frac{\omega_j^2}{\omega_0^2}+m
        \end{align}
       
    $\Theta(t-t')$ represents the Heaviside Theta function \cite{abramowitz1968handbook} and $\mu_d(t-t')$ and $\mu_{od}(t-t')$ represent the diagonal and off-diagonal parts of $\mu$ respectively \cite{bhattacharjee2023decoherence}. The force $\vec{\mathcal{F}}_2$ retains the oscillatory sine term and becomes negligible at long times. Hence the system is driven by the random noise originating primarily from the force $\vec{\mathcal{F}}_1$.
    \begin{align}
\langle &\mathcal{F}_{1\alpha}(t)\rangle=0\\
\frac{1}{2}\langle{\lbrace \mathcal{F}_{1\alpha}(t),\mathcal{F}_{1\beta}(0)}\rbrace\rangle =&\frac{\hbar \delta_{\alpha \beta}}{ 2\pi}\int_{-\infty}^\infty{{d{\omega}}Re[\mu_d({\omega})]}
\frac{{\omega}^3 (m_r+m_b)}{\left(m{\omega}_0^2+m_b \omega^2+K\right)}    \coth\bigg(\frac{\hbar{\omega}}{2k_BT}\bigg) e^{-i{\omega} t} \label{noisecorA}
\end{align}

Here we have considered equal coupling strengths for all the bath oscillators that yields $d_j=d$ and $g_j=g$. In Eq.(\ref{noisecorA}), we choose $d=g=1$ which confirms the equal weightage to the position and momentum coordinate couplings in the system. Further we have taken the approximation that the masses of the bath harmonic oscillators are equal: $m_j = m_b$ for all values of $j$. Within this approximation $K_j$ becomes $j$ independent and is denoted by $K$. This noise correlation function derived in Eq.(\ref{noisecorA}) is used to calculate the decoherence factors in the subsequent sections.
\section{Master equation for a charged particle in a magnetic field}
We set up a master equation for a system coupled to an external heat bath considering the coupling between the system and the bath to be weak and the environment to be sufficiently large for the Born Approximation to hold true \cite{Maximilianbook,Breuer}. Within this framework, the memory-free environment  and the time-local evolution equations give the Born-Markovian Master equation for the total density operator in the interaction picture as \cite{Maximilianbook,Breuer}:
\begin{align}
    \frac{\partial}{\partial t} \rho^{(I)}(t)= \frac{1}{i\hbar}\left[H_{int}(t),\rho^{(I)}(t) \right] \label{Neumann}
\end{align}
where $H_{int}(t)$ represents the time dependent interaction Hamiltonian.\\
Using the Liouville-von Neumann equation (Eq.(\ref{Neumann})), one can derive the general Born-Markov master equation as \cite{Redfield,Blum}:
\begin{align}
    \frac{\partial}{\partial t} \rho_s(t)=-\frac{i}{\hbar}\left[H_s,\rho_s(t) \right]-\frac{1}{\hbar^2}\left \lbrace \left[S,B\rho_s(t) \right]  + \left[\rho_s(t)C,S \right] \right \rbrace \label{Born-Markovgen}
\end{align}
where,
\begin{align}
    B_\alpha=\int_0^\infty d\tau \sum_\beta C_{\alpha \beta}(\tau) S_\beta ^{(I)}(-\tau) \label{Balphagen}\\
    C_\alpha=\int_0^\infty d\tau \sum_\beta C_{\beta \alpha}(-\tau) S_\beta ^{(I)}(-\tau) \label{Calphagen}
\end{align}
The operators with superscript $(I)$ are in the interaction picture and the remaining operators are $\text{Schr\"{o}dinger}$ picture operators and $\tau =t-t^{'}$, which gives a measure of the time through which the environment retains the memory of the interaction with the system. \\
In the non-Markovian limit \cite{Schlosshauer,Caldeira3,Caldeira1,HuPaz,Unruh,Intravaia,Ferialdi} the form of the master equation is the same, however, the operators $B_\alpha$ and $C_\alpha$ are time dependent.
\begin{align}
   B_\alpha(t)=\int_0^t d\tau \sum_\beta C_{\alpha \beta}(\tau) S_\beta ^{(I)}(-\tau)\label{Balphanonmarkov}\\
    C_\alpha(t)=\int_0^t d\tau \sum_\beta C_{\beta \alpha}(-\tau) S_\beta ^{(I)}(-\tau) \label{Calphanonmarkov}
\end{align}
$C_{\alpha \beta}$ and $C_{\beta \alpha}$ representing the environmental self correlation functions and $S_{\beta}$ is the system operator \cite{bhattacharjee2023decoherence}.\\
It is to be noted that in the non-Markovian derivation of the master equation, the Born Approximation is assumed to hold true (weak coupling limit) \cite{Maximilianbook}.
\section{Decoherence in the presence of a magnetic field and a harmonic oscillator potential}
\begin{figure}[H]
\centering
\hspace*{-0.8cm}\includegraphics[scale=0.7]{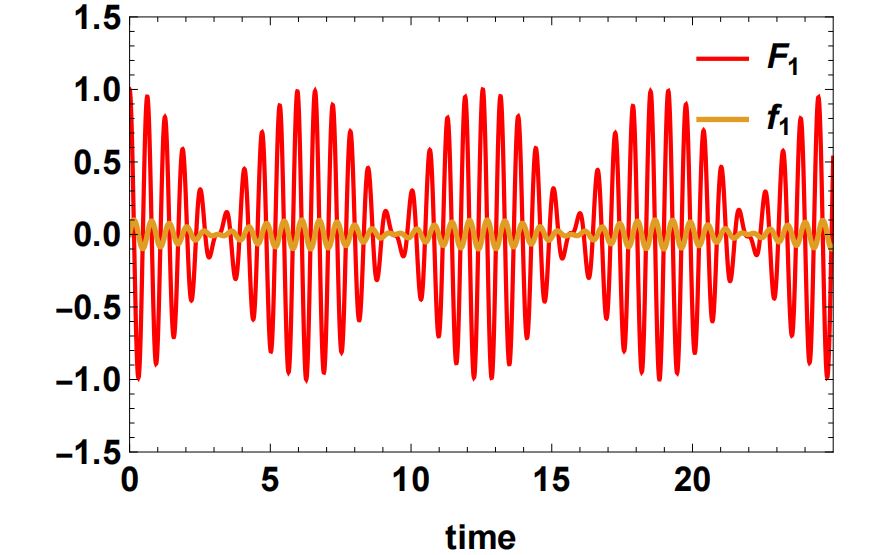} 
\caption{(a) $F_1$, 
$f_1$ versus time for $\omega_0=10$ and $\omega_c=1$}
\label{schematic}
\end{figure}

In this section, we derive the Born Markovian master equation for an open quantum system corresponding to a harmonically oscillating charged Brownian particle in the presence of an external magnetic field. The Hamiltonian in Eq.(\ref{Hamiltonian}) can be modelled in the form of a system environment (bath) interaction Hamiltonian as:
\begin{align}
    H = H_S + H_E + H_{SE}
\end{align}
    where,
    \begin{align}
    H_{S} &= \frac{1}{2m}\left(\vec{p}-\frac{q\vec{A}}{c}\right)^{2} +
    \frac{1}{2} m\omega_0^{2}(x^{2}+y^{2})\label{syshamiltonian}\\
     H_{E}&=\sum_{i} \frac{p_{i}^{2}}{2m_{i}} +\frac{1}{2}m_{i}\omega_{i}^{2}q_{i}^{2} \label{envhamiltonian}
\end{align}
and the interaction Hamiltonian $H_{SE}$ is modeled as \cite{bhattacharjee2023decoherence}:
\begin{equation}
    H_{SE} = x \otimes \sum_{i} q_{ix} + y \otimes \sum_{i} q_{iy}+ p_x \otimes \sum_{i} p_{ix} + p_y \otimes \sum_{i} p_{iy}\label{sys-envhamiltonian}
\end{equation}
$x$, $y$, $p_x$, $p_y$ represent the position and the momentum coordinates of the Brownian particle and $q_{ix}$, $q_{iy}$, $p_{ix}$, $p_{iy}$ represent the position and momentum coordinates of the $i^{th}$ bath oscillator respectively.   \\
Now we solve the equations of motion for the system Hamiltonian (Eq.(\ref{syshamiltonian})) and arrive at the following expressions for the position coordinates of the Brownian particle in the interaction picture:
\begin{align}
    x(\tau)=&\frac{1}{2\sqrt{4 \omega_0^2+\omega_c^2}}\bigg[\bigg \lbrace\left(-\omega_c+\sqrt{4 \omega_0^2+\omega_c^2} \right) \cosh(A\tau)+ \notag \\
    &\left(\omega_c+\sqrt{4\omega_0^2+\omega_c^2} \right) \cosh(B\tau) \bigg \rbrace X +\bigg \lbrace 2 \omega_0^2  \left(\frac{\sinh(A \tau)}{A}-\frac{\sinh(B \tau)}{B} \right) \bigg \rbrace Y+\notag \\
   &\bigg \lbrace \left(\omega_c+ \sqrt{4 \omega_0^2+\omega_c^2} \right)\frac{\sinh(A \tau)}{m A}+ \left(-\omega_c+\sqrt{4\omega_0^2+\omega_c^2} \right)\frac {\sinh(B \tau)}{m B} \bigg \rbrace P_x +\notag \\ &\bigg \lbrace \frac{2}{m} \left( -\cosh(A \tau)+\cosh(B \tau)\right) \bigg \rbrace P_y\bigg] \label{couplx}\\
   y(\tau)=&\frac{1}{2\sqrt{4 \omega_0^2+\omega_c^2}}\bigg[\bigg \lbrace 2 \omega_0^2  \left(\frac{\sinh(B \tau)}{A}-\frac{\sinh(A \tau)}{B} \right) \bigg \rbrace X-\notag \\
   &\bigg \lbrace \left(-\omega_c+\sqrt{4 \omega_0^2+\omega_c^2} \right) \cosh(A\tau)+\left(\omega_c+\sqrt{4\omega_0^2+\omega_c^2} \right) \cosh(B\tau) \bigg \rbrace Y+\notag \\
   &\bigg \lbrace \frac{2}{m} \left(\cosh(A \tau)-\cosh(B \tau)\right) \bigg \rbrace P_x+\bigg \lbrace \left(\omega_c+ \sqrt{4 \omega_0^2+\omega_c^2} \right)\frac{\sinh(A \tau)}{m A}+ \notag \\
   &\left(-\omega_c+\sqrt{4\omega_0^2+\omega_c^2} \right)\frac {\sinh(B \tau)}{m B} \bigg \rbrace P_y\bigg]\label{couply}
   \end{align}
 where $\omega_c=  
 qB/m$ is the cyclotron frequency and
   \begin{align}
       &A=\frac{\sqrt{-2 \omega_0^2-\omega_c^2-\omega_c\sqrt{4 \omega_0^2+\omega_c^2}}}{\sqrt{2}} \\
      & B=\frac{\sqrt{-2 \omega_0^2-\omega_c^2+\omega_c\sqrt{4 \omega_0^2+\omega_c^2}}}{\sqrt{2}}\\
       &X=x(0), Y=y(0), P_x=m\Dot{x}(0), P_y=m\Dot{y}(0)
   \end{align}
    $X$ and $Y$ are the position operators in the $\text{Schr\"{o}dinger}$ picture and $P_x$, $P_y$ are the momentum in the $x$ and $y$ directions respectively, representing the $\text{Schr\"{o}dinger}$ picture momentum operators.\\
   Using Eqs.(\ref{couplx}) and (\ref{couply}) for the system operators in the generalized master equation Eq.(\ref{Born-Markovgen}) and retaining only the decoherence terms, one gets:
    \begin{align}
       \frac{\partial \rho_s}{\partial t}=&-\frac{1}{\hbar}\int_0^ t d\tau \nu(\tau)F_1(\tau)\left[X,\left[X,\rho_s(t) \right]  \right]-\frac{1}{\hbar}\int_0^t d\tau \nu(\tau)F_1(\tau)\left[Y,\left[Y,\rho_s(t) \right]  \right]-\notag \\
      & \frac{1}{\hbar}\int_0^t d\tau \nu(\tau)F_2(\tau)\left[X,\left[Y,\rho_s(t) \right]  \right]+\frac{1}{\hbar}\int_0^t d\tau \nu(\tau)F_2(\tau)\left[Y,\left[X,\rho_s(t) \right]  \right]- \notag \\
      &\frac{1}{\hbar}\int_0^t d\tau \nu(\tau)f_1(\tau)\left[X,\left[P_x,\rho_s(t) \right]  \right]-\frac{1}{\hbar}\int_0^t d\tau \nu(\tau)f_1(\tau)\left[Y,\left[P_y,\rho_s(t) \right]  \right]-\notag \\
       &\frac{1}{\hbar}\int_0^t d\tau \nu(\tau)f_2(\tau)\left[X,\left[P_y,\rho_s(t) \right]  \right]+\frac{1}{\hbar}\int_0^t d\tau \nu(\tau)f_2(\tau)\left[Y,\left[P_x,\rho_s(t) \right]  \right] \label{decequation}
   \end{align}
   $\nu(\tau)$ in the master equation represents the random noise kernel that captures the effect of the environment (heat bath) on the system. It is dependent on the spectral density of the environmental oscillators and can be obtained from Eq.(\ref{noisecorA}), derived from the Quantum Langevin dynamics \cite{bhattacharjee2023decoherence,malay,Li}:
   \begin{align}
   \nu(\tau)=\int_{0}^\infty{{d{\omega}}J(\omega)}
\frac{{\omega}^2 (m_r+m_b)}{\left(m{\omega}_0^2+m_b \omega^2+K\right)}    \coth\bigg(\frac{{\omega}}{\Omega}\bigg) \cos(\omega \tau) \label{noisekernel}
\end{align}
   $J(\omega)$ is the spectral density of the bath and $\Omega$ is the temperature dependent thermal frequency given by $\Omega=\frac{2 k_B T}{\hbar}$ \cite{bhattacharjee2022pramana}. Here we present an analysis for the Ohmic bath model where the spectral density varies linearly with frequency $\omega$ and the proportionality constant $\gamma$ represents the damping coefficient for the Brownian system. 
  Here one upper cut-off frequency $\Lambda$ has been chosen such that above this value the density of the bath oscillators abruptly goes to zero.
   \begin{equation}
J(\omega)=
    \begin{cases}
      \gamma \omega, & \text{for}\ \omega<\Lambda \\
      0, & \hspace{0.6cm}\omega \geq\Lambda
    \end{cases}
  \end{equation}
So, the time evolution equation for the density matrix $\rho_s$ (Eq.(\ref{decequation})) can be written as: 
\begin{align}
       \frac{\partial \rho_s}{\partial t}=&-D_1\left[X,\left[X,\rho_s(t) \right]  \right]-D_1\left[Y,\left[Y,\rho_s(t) \right]  \right]-D_2\left[X,\left[Y,\rho_s(t) \right]  \right]+D_2\left[Y,\left[X,\rho_s(t) \right]  \right]- \notag \\
      &\mathscrsfs{D}_1\left[X,\left[P_x,\rho_s(t) \right]  \right]-\mathscrsfs{D}_1\left[Y,\left[P_y,\rho_s(t) \right]  \right]-\mathscrsfs{D}_2\left[X,\left[P_y,\rho_s(t) \right]  \right]+\mathscrsfs{D}_2\left[Y,\left[P_x,\rho_s(t) \right]  \right] \label{decequation1}
         \end{align}
  where, we have considered $\hbar=1$ and the decoherence factors $D_{1(2)}$ and $\mathscrsfs{D}_{1(2)}$ are:
\begin{align}
    D_{1(2)}(t)=\int_0^t d\tau \nu(\tau) F_{1(2)} (\tau) \label{D1}\\
    \mathscrsfs{D}_{1(2)}(t)=\int_0^t d\tau \nu(\tau) f_{1(2)} (\tau)
\end{align}
and
\begin{align}
       F_1(\tau)=&\frac{\left(-\omega_c+\sqrt{4\omega_0^2+\omega_c^2} \right)\cosh(A \tau)+\left(\omega_c+\sqrt{4\omega_0^2+\omega_c^2} \right)\cosh(B \tau)}{2 \sqrt{4\omega_0^2+\omega_c^2}} \label{F1}\\
   &F_2(\tau)=\frac{\omega_0^2 \left(B \sinh(A \tau)-A\sinh(B \tau) \right)}{AB\sqrt{4\omega_0^2+\omega_c^2}} \label{F2}\\
&f_1(\tau)=\frac{\left(\omega_c+\sqrt{4\omega_0^2+\omega_c^2} \right) B\sinh(A \tau)+\left(-\omega_c+\sqrt{4 \omega_0^2+\omega_c^2} \right) A \sinh(B \tau)}{ 2 A B m \sqrt{4\omega_0^2+\omega_c^2}} \label{f1}\\
&f_2(\tau)=\frac{-\cosh(A \tau)+\cosh(B \tau)}{m \sqrt{4 \omega_0^2+\omega_c^2}} \label{f2}
   \end{align} 
It is to be noted that in Eq.(\ref{decequation1}) the first four terms represent the Lindblad double commutator form of spatial decoherence, whereas the last four terms denote decoherence originating from anomalous diffusion \cite{Schlosshauer,bhattacharjee2023decoherence}. The contributions of the anomalous diffusion terms are negligible in comparison to the terms involving $D_1$ and $D_2$, at long times, even in the low temperature regime \cite{Sinhadecoherence}.
Further, the Lindblad double commutators involving the x and y cross terms cancel one another and hence the decoherence in this system is primarily governed by the first and second Lindblad double commutator terms on the right in Eq.(\ref{decequation1}) and these depend on the decoherence factor $D_1$.\\
$F_1$ and $f_1$ are plotted in Fig.(\ref{schematic}) which proves our statement regarding $F_1$ dominating over $f_1$ throughout the entire time range, leading to the dominance of the  decoherence factor $D_1$ over $\mathscrsfs{D}_1$ in the decay of the off-diagonal terms of the density matrix.
Using Eq.(\ref{D1}) in Eq.(\ref{decequation1}), one arrives at the following form for the decay of the off-diagonal elements of the density matrix :
\begin{equation}
 \rho_s(x,x',y,y',t)=\rho_s(x,x',y,y',0) \exp \left[- D(t)\right]
 \label{densitydecay}\\
\end{equation}
where $D(t)$ is given by:
\begin{equation}
 D(t)= \left \lbrace \left( \Delta x\right)^2 +
 \left(\Delta y\right)^2 \right \rbrace\int_0^t D_1(t')dt'\label{decorate}
\end{equation}
Here, $\Delta x=x-x^{'}$, $\Delta y=y-y^{'} $ and 
\begin{align}
    D_{1}(t)=\int_0^t d\tau \nu(\tau) F_{1} (\tau) 
\end{align}

\section{Results and Discussions}
\begin{figure}[H]
\centering
\hspace*{-1cm}\includegraphics[scale=1.3]{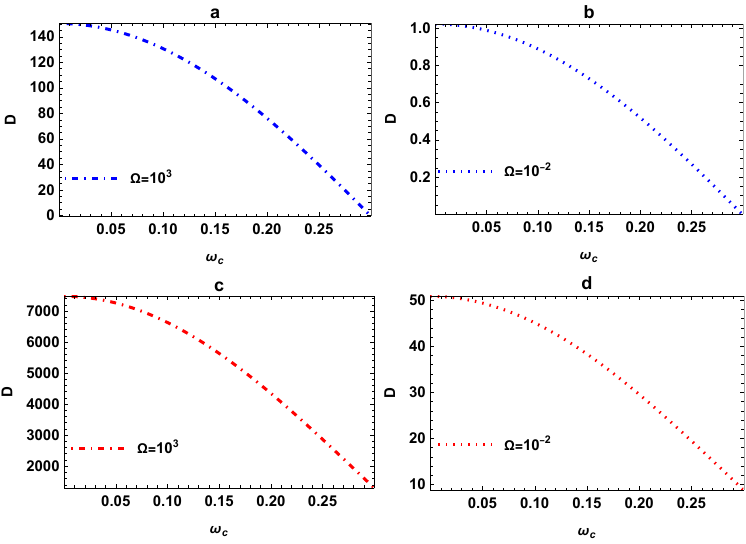} 
\caption{Decoherence factor D(t) (at t=10) versus $\omega_c$ with $m=1$, $K=10^2$, $m_b=10^{-2}$, $m_r=10^{-3}$, $\Lambda=10^3$, $\Delta x=\Delta y=1$: (a) High temperature ($\Omega=10^{3}$), $\omega_0=10$, $\gamma=1$; (b) Low temperature ($\Omega=10^{-2}$), $\omega_0=10$, $\gamma=1$; (c) High temperature ($\Omega=10^{3}$), $\omega_0=1$, $\gamma=10$; (d)  Low temperature ($\Omega=10^{-2}$), $\omega_0=1$, $\gamma=10$.} 
\label{decoherence}
\end{figure}

\begin{figure}[H]
\centering
\hspace{-0.5cm}\includegraphics[scale=1.1]{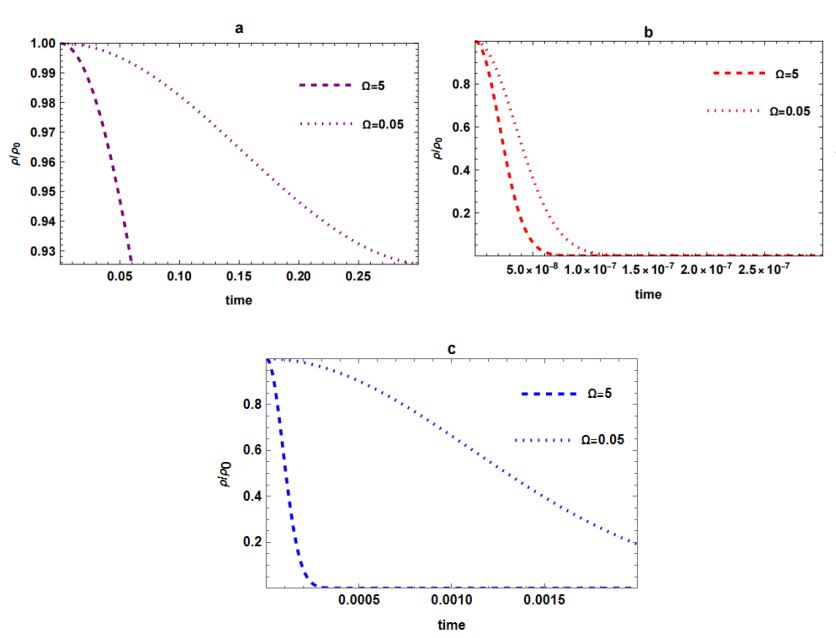} 
\caption{$\rho/\rho_0$ versus time for the (a) position coupling ($d=1$, $g=0$, $K=0$); (b) momentum coupling ($d=0$, $g=1$, $K=10^2$) and (c) position and momentum coupling  ($d=1$, $g=1$, $K=10^2$) cases with $m=1$, $\omega_0=10$, $\omega_c=1$, $m_b=10^{-2}$, $m_r=10^{-3}$, $\Lambda=10^3$, $\Delta x=\Delta y=1$.} 
\label{decoherencecomp}
\end{figure}

In Fig.(\ref{decoherence}), we have displayed the magnetic field dependence of the decoherence factor $D(t)$ at high and low temperatures for different values of the damping parameter $\gamma$ and harmonic frequency $\omega_0$. Fig.(\ref{decoherencecomp}) shows a comparison of decoherence in the presence of both position and momentum coordinate couplings and decoherence in the presence of only position coordinate and momentum coordinate couplings separately.
Notice that in Fig.(\ref{decoherence}), an increase in the cyclotron frequency (due to an increase in the applied magnetic field) leads to a reduction in the decoherence factor $D$ at all temperature regimes. The reduction in the decoherence factor leads to a slower fall off of the off-diagonal elements of the reduced density matrix in the open quantum system \cite{bhattacharjee2023decoherence}.
This is consistent with the observation that the cyclotron frequency associated with a magnetic field leads to coherent
oscillations , which play a vital role in delaying the loss of information to the environment, as discussed in several earlier papers \cite{Ghorashi,Tcoffo,Germain,Armel,bhattacharjee2023decoherence}. 
At a given temperature, one notices that the under-damped case ($\omega_0>\gamma$) (Fig.(\ref{decoherence}a,c)) displays a stronger dependence on the cyclotron frequency compared to the over-damped case ($\gamma>\omega_0$)  (Fig.(\ref{decoherence}b,d)), 
which is consistent with our expectation that the oscillatory effects of the cyclotron frequency  
play a crucial role in shaping the dynamics of the system at low damping.\\
In Fig.(\ref{decoherencecomp}), we have plotted $\rho/\rho_0$, the off-diagonal element of the reduced density matrix as a function of time, 
which measures the extent of decoherence stemming from interaction with the environment. The Figures show that the rate of decoherence is very low in Fig.(\ref{decoherencecomp}a), where the the system and the bath are coupled only through position coordinates and the loss of coherence is much faster in Fig.(\ref{decoherencecomp}b), pertaining to a momentum coordinate system-bath coupling. Fig.(\ref{decoherencecomp}c) which corresponds to a case where both position and momentum couplings are present, exhibits a rate of decoherence between these two extreme cases (position coupling and momentum coupling).
It therefore follows that position coupling slows down the loss of information from the system to the environment and the two coupling parameters can be tuned to get the desired rate of decoherence in real physical systems. 
It is evident from all the plots that a rise in temperature leads to a faster decay of $\rho/\rho_0$, resulting in a faster Quantum-to-Classical transition in the high temperature domain, consistent with earlier studies \cite{Caldeira1}.
\section{Conclusion}
In this paper, we have considered the Hamiltonian for an open quantum system where a harmonically oscillating charged Brownian particle is coupled to an Ohmic heat bath via both position and momentum coordinates, in the presence of a magnetic field. We have derived the corresponding quantum Langevin equation for the system following the Ford, Lewis and O' Connell' (FLO) approach \cite{Li}.
In the subsequent sections, we have formulated the non-Markovian master equation involving normal and anomalous diffusion terms.
We study the temporal decay of the off-diagonal elements of the reduced density matrix for the Brownian motion model which corresponds to the destruction of  superposition of states leading to a Quantum-to-Classical transition. It has been shown analytically and numerically (see Fig.(\ref{schematic})) that the contributions of the anomalous diffusion term $\mathscrsfs{D}_1$ is negligible in comparison to the normal decoherence factor $D_1$. The process of decoherence
is faster in the high temperature regime compared to the low temperature regime as quantum noise correlations slow down the process of loss of coherence
in the quantum domain \cite{Sinhadecoherence,bhattacharjee2023decoherence}. 
Moreover, the presence of a cyclotron frequency, originating from the applied magnetic field, results in an oscillatory behaviour of the Brownian particle and consequently the loss of information is delayed, as seen in Fig.(\ref{decoherence}).
In one of our earlier works we have shown that decoherence occurs at a slower rate in the presence of a high magnetic field \cite{bhattacharjee2023decoherence}.
Similar effects are seen in the context of a single spin associated with a nitrogen vacancy defect in diamond, where one notices that the magnetic field has a confining effect on the system and delays the onset of decoherence \cite{jamonneau}.
Furthermore, the loss of information from the system occurs at a higher rate when there are multiple channels of interaction between the system and the environment. 
We have shown that decoherence occurs at a slower rate for a Brownian particle coupled to a heat bath through position coordinates only, which is the case for most of the earlier studies on decoherence \cite{Ghorashi,Tcoffo,Germain,Armel}. However in the present analysis we have considered the coupling between the system and the environment via both position and momentum coordinates. In some earlier works, the Langevin equation for a Brownian particle was derived for the momentum coupling case and the noise correlation was seen to be drastically different from the position coupling case \cite{malay,bhattacharjee2022quantum}. Similarly the inclusion of both position and momentum couplings induces a variation in the noise kernel (Eq.\ref{noisekernel}) that results in the change in the rate of decoherence (see Fig.(\ref{decoherencecomp})). The loss of information occurs at a faster rate in the presence of momentum coupling, as the effect of environment-induced random noise is more pronounced when the momentum of the Brownian particle and the bath oscillators are coupled. However, the off-diagonal terms of the reduced density matrix in the presence of both position and momentum coupling exhibit a rate of decay that lies between the decay rates pertaining to the position coupling and the momentum coupling cases. Thus it can be inferred that 
the position coupling lowers the rate of decay of $\rho/\rho_0$, whereas, the momentum coupling enhances the loss of information from the system. This result is it agreement with that of \cite{Ferialdi}, where the authors have shown that the effect of dissipation is more pronounced in the presence of momentum coupling which leads to an accelerated relaxation to equilibrium and a faster decoherence in an anomalous diffusive system.  Thus our analysis 
indicates a way to modulate the rate of decoherence in a controlled manner by tuning the position and the momentum coupling parameters `$d$' and `$g$' respectively. Such a system involving anomalous momentum coupling can be physically realized in the study of effect of the electromagnetic black body radiation on Josephson junctions \cite{cuccoli,Leggetttunneling,kohler1,kohler2}. 
In several real physical systems, both momentum and position coordinates contribute to the environmental couplings and are analyzed in detail to control the two channels of decay of the off-diagonal terms of the reduced density matrix \cite{kohler1,kohler2}. Moreover, the results can also be utilized to investigate the non-Markovian dynamics of  molecular compounds, mesoscopic islands coupled to fluctuating charges, and the transport of charged particles moving under the influence of random magnetic
fields \cite{Ankerhold}. \\
Decoherence was first experimentally tracked
by creating a mesoscopic superposition of quantum states involving radiation fields with classically distinct phases and its progressive decoherence was observed. The experiment involved Rydberg atoms interacting one at a time with a few photon coherent field trapped in a high $Q$ microwave cavity \cite{Brune}. 
However, some modern experimental techniques have gone beyond this study and investigated decoherence in the presence of a large number of atoms in a background gas resulting in a suppression of interference with increasing pressure of the gas \cite{hackermuller}. The decoherence of a gas of strongly interacting bosons in an optical lattice exposed to near-resonant light and spontaneous emission was studied in \cite{bouganne}. Our results for the study of decoherence in the presence of both position and momentum coordinate couplings are of current relevance and 
can be tested by designing a suitable cold atom-ion experimental setup\cite{coldiondecexpt}. 

\section{Acknowledgement}
SB acknowledges Raman Research Institute where the problem was initially formulated.
%
\end{document}